\documentclass[%
 reprint,
 amsmath,amssymb,
 aps,pra
]{revtex4-2}

\usepackage{booktabs,array}
\usepackage{graphicx}% Include figure files
\usepackage{dcolumn}% Align table columns on decimal point
\usepackage{bm}
\usepackage{amsmath,mathtools}
\usepackage{pgfplots}
\usepackage{xcolor}
\usepgfplotslibrary{fillbetween}
\usepgfplotslibrary{patchplots}
\usepackage{multirow}
\usepackage{gensymb}
\usepackage{comment}
\usepackage[normalem]{ulem}
\usepackage{tikz}
\usetikzlibrary{shapes.geometric,arrows.meta,positioning}

\definecolor{mygreen}{rgb}{0.1, 0.75, 0.1}
\definecolor{refcolor}{rgb}{0.0, 0.0, 0.0}
\definecolor{myscan}{rgb}{0.0, 0.0, 1.0}
\definecolor{mym062x}{rgb}{1.0, 0.0, 0.0}
\definecolor{myhf}{rgb}{0.0, 0.6, 0.0}

\begin{document}

%\preprint{APS/123-QED}

\title{Comparison of Density-Matrix Corrections to Density Functional Theory}

\author{Daniel Gibney}
\author{Jan-Niklas Boyn}%
\author{David A. Mazziotti}%
\email{damazz@uchicago.edu}%
\affiliation{Department of Chemistry and The James Franck Institute, The University of Chicago, Chicago, Illinois 60637 USA}%
\date{Submitted June 13, 2022\textcolor{black}{; Revised October 7, 2022}}% It is always \today, today,

\begin{abstract}
%\section*{Abstract}
Density functional theory (DFT), one of the most widely utilized methods available to computational chemistry, fails to describe systems with statically correlated electrons. To address this shortcoming, in previous work we transformed DFT into a one-electron reduced density matrix theory (1-RDMFT) via the inclusion of a quadratic one-electron reduced density matrix (1-RDM) correction. Here, we combine our 1-RDMFT approach with different DFT functionals as well as Hartree-Fock to elucidate the method's dependence on the underlying functional selection. Furthermore, we generalize the information density matrix functional theory (iDMFT), recently developed as a correction to the Hartree-Fock method, by incorporating density functionals in place of the Hartree-Fock functional. We relate iDMFT mathematically to our approach and benchmark the two with a common set of functionals and systems.\\

%Optimizing our adjustable parameter, $w$, independently for each system studied, we gain insight into the magnitude of the static correlation correction required for different functionals.

\end{abstract}

\maketitle

%\section*{TOC Graphic}
%\begin{figure}[h]
%\centering
%\includegraphics[height = 5cm, width = 5cm]{toc.png}
%\label{fig:toc}
%\end{figure}

%\tableofcontents

\section{Introduction}

While density functional theory in the Kohn-Sham formulation (KS-DFT) has seen significant success throughout chemistry due to its computational affordability, as well as its relative accuracy~\cite{scaling1,scaling2,citationstats,ReactionBarriers,vibrations,tddft,ionization,Geometries}, it has been noted to struggle with three significant failures. Namely: i) the self-interaction error, ii) the charge transfer error, and iii) the static correlation error arising from near-degenerate electronic states~\cite{3errors,SIE,disp}. While KS-DFT is, in theory, exact when using the unknown universal functional~\cite{Universalfunctional}, the aforementioned errors are rooted in the approximate nature of current density functionals. Furthermore, in contrast to wavefunction-based theories where there exists a clear path to improving the prediction of electronic properties via the inclusion of higher-order excitations from the Hartree-Fock (HF) reference, e.g. using the configuration interaction (CI) or coupled cluster (CC) approaches, DFT does not offer such a clear systematic path of improvement~\cite{FCI1,FCI2,CC}.\\

Modern functional developments have utilized a wide variety of different approximations with varying complexity to try to improve DFT, resulting in the so-called ``functional zoo.'' A Jacob's ladder style scheme has been conceptualized to aid in functional classification~\cite{Jacobs_ladder}, with the idea being that ascending the ladder to more computationally complex functionals will yield improvements. While modern functionals have generally predicted chemical properties with increasing accuracy ~\cite{Head-Gordon2018}, it has been argued by Medvedev et. al.~\cite{medvedev} and Brorsen et. al.~\cite{Hammes-Schiffer2017} that, while properties predicted by newer functionals are more accurate, their underlying electronic densities are increasingly deviating from the exact density. This disconnect has been attributed to newer density functional approximations relying on fitting to reproduce specific chemical properties of interest from reference calculations or experimental results instead of attempting to improve upon the fundamental quality of DFT, the electronic density~\cite{Truhlar2006,Truhlar2016}. This is due, in part, to the difficulty in identifying and reproducing properties of the universal functional  as compared to simply optimizing a set of parameters to reproduce reference data. Therefore, approaches outside of functional development may be necessary to further improve upon DFT. \\

Several methods have been developed aiming to enable DFT to describe static correlation. These approaches include the expansion of DFT into the complex plane to allow for static correlation to be captured using fractional orbital occupations, which requires transforming the real-valued functionals into the complex plane as well~\cite{MHG}. Complex orbital DFT has recently been expanded to utilize hypercomplex numbers for describing statically correlated systems beyond biradicals.\cite{Su2021} Another approach relies on enforcing the Perdew–Parr–Levy–Balduz (PPLB) flat-plane conditions through a scaling correction~\cite{PPLB,SC1,SC2,SC3}, aiming to recover the piece-wise linearity of density functionals between integer numbers of electrons. These methodologies both utilize fractional occupations as part of their improvement over KS-DFT. \\

Another area of research which focuses on fractional occupations for describing static correlation is presented by one-electron reduced density matrix functional theory (1-RDMFT)~\cite{Gilbert,Schmidt_2019,Buchholz_2019,RDMFT, Piris2021,Schilling_2019}. These approaches utilize Gilbert's theorem to express the ground-state energy as a functional of the one-electron reduced density matrix (1-RDM) \textcolor{black}{rather than the wavefunction $\Psi(12 \hdots N)$~\cite{Gilbert} where
\begin{equation}
    ^{1} D(1;\bar{1}) = \int{ \Psi(12 \hdots N) \Psi^{*}({\bar 1}2 \hdots N) d2 \hdots dN }
\end{equation}
\noindent with each roman number representing the spatial and spin coordinates of an electron.}  Using the 1-RDM as the fundamental variable allows for the description of static correlation through fractional occupation numbers~\cite{Schmidt_2019,Buchholz_2019,RDMFT}. Natural orbital functional theory (NOFT) presents a related approach which uses the natural orbitals (NOs) and their occupation numbers (NONs), which may be obtained from the 1-RDM, to reconstruct the 2-RDM subject to $N$-representability conditions~\cite{PNOF1,PNOF2,PNOF3,Piris2017,Piris2021}. Taking inspiration from these methods we previously transformed KS-DFT into a 1-RDMFT to retain the favorable computational scaling of KS-DFT while enabling the description of static correlation~\cite{SDP_paper, Gibney2022}.\\

In this article we further expand upon the theory of translating DFT into a 1-RDMFT framework and compare how our current implementation of our 1-RDMFT method relates to the iDMFT method developed by Wang et. al~\cite{iDMFT,Schilling2022}. To facilitate this comparison, we generalize iDMFT to use density functionals in addition to the Hartree-Fock functional. Since our 1-RDMFT approach, as well as iDMFT rely on functional selection, we perform benchmarking calculations on a test set of small statically correlated molecules to elucidate the functional dependence of the two surveyed methods. Finally, the magnitude of the correction term required with a given functional for 1-RDMFT or iDMFT, obtained from our benchmark, provides insight into its inherent ability to describe multi-reference correlation. \\

\section{Theory}

We review and expand upon the conversion of DFT into 1-RDMFT in Section~\ref{sec:convert} and compare the resulting 1-RDMFT with iDMFT in Section~\ref{sec:cnnect}.

\subsection{Conversion of DFT into a 1-RDMFT}  \label{sec:convert}

Consider the energy functional for DFT
\begin{equation}
\label{eq:DFT}
    E_{\rm DFT}[\rho] = T_{\rm s}[\rho] + V[\rho] + F_{\rm xc}[\rho] ,
\end{equation}
where $\rho$ is the one-electron density, $T_{\rm s}[\rho]$ is the non-interacting kinetic energy functional---the kinetic energy from the single Slater determinant that yields the density $\rho$, $V[\rho]$ is the sum of the one-electron (external) potential and the Coulomb potential, and $F_{\rm xc}[\rho]$ is the exchange-correlation functional.  \textcolor{black}{As in Ref.~\cite{Gibney2022}, we convert DFT into a 1-RDMFT by replacing the non-interacting kinetic energy by the full kinetic energy and adding a 1-RDM based correction functional $C[^{1} D]$
\begin{equation}
    E_{\rm RDMFT}[^{1} D] = E_{\rm DFT+T}[^{1} D]+ C[^{1} D] ,
\end{equation}
where $E_{\rm DFT+T}[^{1} D]$ is defined as }
\begin{equation}
E_{\rm DFT+T}[^{1} D] = E_{\rm DFT}[\rho] + (T[^{1} D]-T_{\rm s}[\rho]) .
\end{equation}
\noindent Because the exchange-correlation functionals in traditional DFT are not exact, we can treat $C[^{1} D]$ as a general functional that\textcolor{black}{, as part of its conversion of DFT into a 1-RDMFT,} also accounts for the limitations of existing functionals to treat static electron correlation.  \\

\textcolor{black}{The correction has several advantages relative to traditional approaches in DFT and 1-RDMFT.  From the perspective of 1-RDMFT, the correction allows us to build upon the wealth of functionals that have been developed for DFT as well as the low computational scaling afforded by DFT's exchange-correlation potential.  From the perspective of DFT, the correction allows us to use explicit 1-RDM information for an improved treatment of static correlation.} \\

\textcolor{black}{As a practical correction to DFT, we focus on approximating the part of the correction $C[\prescript{1}{}{D}]$ that lowers the energy from the presence of static electron correlation.  We assume that this part of the correction functional:} i) obeys particle-hole symmetry, ii) vanishes in the limit of no correlation, and iii) rewards the formation of fractional occupation for orbitals as they near energetic degeneracy.  Note that assumption ii) is an approximation for the exact $C[\prescript{1}{}{D}]$ functional.  Using these assumptions, we previously obtained the following form in Ref.~\cite{Gibney2022}:
\begin{equation} \label{eq:Original_correction}
    E_{\text{RDMFT}}[\prescript{1}{}{D}]=E_{\text{DFT+T}}[\prescript{1}{}{D}]-\text{Tr}[(\prescript{1}{}{W}\textcolor{black}{(\prescript{1}{}{D}-\prescript{1}{}{D}^{2})}],
\end{equation}
where $\prescript{1}{}{W}$ is an arbitrary positive semidefinite weight matrix. By taking $\prescript{1}{}{W}$ to be a weighted identity matrix $w\prescript{1}{}{I}$, we produce the final form of our correction~\cite{Gibney2022}:
\begin{equation} \label{eq:fin_correction}
    E_{\text{RDMFT}}[\prescript{1}{}{D}]=E_{\text{DFT+T}}[\prescript{1}{}{D}]+w(\text{Tr}[\prescript{1}{}{D}^{2}-\prescript{1}{}{D}])\,.
\end{equation}
If the 1-RDM is idempotent, we note that this correction vanishes, and if the 1-RDM is not idempotent, it is nonzero and serves to remove the double counting of the correlated kinetic energy and to account for static correlation that is missing from traditional DFT. \\

When $w=0$, the energy functional is readily minimized by a conventional Kohn-Sham self-consistent-field (SCF) calculation; when $w \neq 0$, the functional can be minimized by a modified Kohn-Sham SCF calculation where the modified Kohn-Sham energy is given by
\begin{equation}
\label{eq:MKS1}
 E_{\rm MKS} = \text{Tr}[ H_{\rm KS} \, ^{1} D] + w(\text{Tr}[\prescript{1}{}{D}^{2}-\prescript{1}{}{D}]) .
\end{equation}
To express the minimization of $E_{\rm MKS}$ at each SCF iteration as a semidefinite program (SDP), we can relax the quadratic term by introducing an auxiliary matrix variable $^{1} F$
\begin{equation}
\label{eq:MKS2}
 E_{\rm MKS} = \text{Tr}[ H_{\rm KS} \, ^{1} D] + w(\text{Tr}[\prescript{1}{}{F}-\prescript{1}{}{D}]) ,
\end{equation}
in which
\begin{equation}
\left(\begin{array}{cc}
^{1} I & ^{1} D
\\
 ^{1} D & ^{1} F
\end{array}\right) \succeq 0 .
\end{equation}
The semidefinite constraint causes the trace of $^{1} F$ to be bounded by the trace of the 1-RDM squared~\cite{Gibney2022}, and hence, the minimization of Eq.~(\ref{eq:MKS2}) as an SDP is equivalent to the minimization of Eq.~(\ref{eq:MKS1}). \textcolor{black}{ The overall algorithm for our 1-RDMFT procedure is summarized in Fig.~\ref{fig:flow}}. It should be noted that the modified Kohn-Sham energy, just like the Kohn-Sham energy in DFT, does not yield the energy of the system, which instead is obtained by evaluating Eq.~(\ref{eq:fin_correction}) using the converged 1-RDM. \\

%\begin{comment}
\tikzstyle{startstop} = [rectangle, minimum width=0cm, minimum height=0cm,text centered, draw=black, fill=gray!30]
\tikzstyle{io} = [rectangle, minimum width=2cm, minimum height=1cm, text centered, draw=black, fill=red!40]
\tikzstyle{process} = [rectangle, minimum width=0cm, minimum height=0cm, align = left, draw=black, fill=blue!30]
\tikzstyle{decision} = [rectangle, minimum width=0cm, minimum height=0cm, text centered, draw=black, fill=red!40]
\tikzstyle{arrow} = [thick,->,>=stealth]

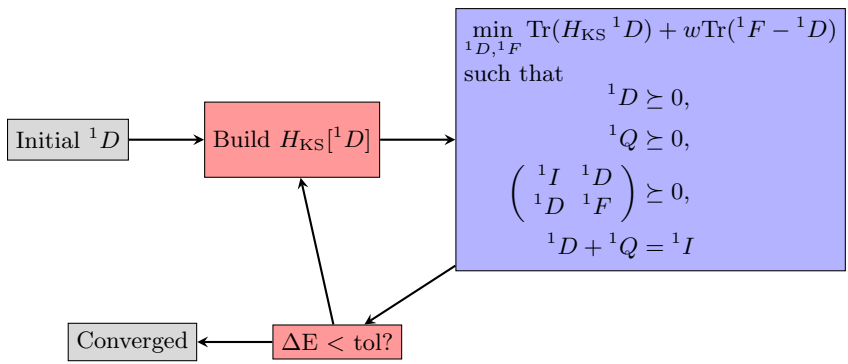
\begin{figure*}

\begin{tikzpicture}[node distance=1cm]
\node (start) [startstop] {Initial $\prescript{1}{}{D}$};
\node (fock) [io, right=of start] {Build $H_{\text{KS}}[\prescript{1}{}{D}]$};
%\begin{comment}
%\min\limits_{x_1,\dots x_T}
\node (sdp) [process, right=of fock] {$\min\limits_{\prescript{1}{}{D},\prescript{1}{}{F}} \text{Tr}(H_{\text{KS}}\prescript{1}{}{D})+w\text{Tr}(\prescript{1}{}{F}-\prescript{1}{}{D})$\\
such that\\
$\begin{aligned}
~~~~~~~~~~~~~~~~~\prescript{1}{}{D} &\succeq  0, \\
\prescript{1}{}{Q} &\succeq  0, \\
\left(\begin{array}{cc}
^{1} I & ^{1} D
\\
 ^{1} D & ^{1} F
\end{array}\right)
&\succeq  0, \\
\prescript{1}{}{D} + \prescript{1}{}{Q} &= \prescript{1}{}{I} \\
\end{aligned}$
};
%\end{comment}
\begin{comment}
\node (sdp) [process, right=of fock] {$\min_{\prescript{1}{}{D}} \text{Tr}(\prescript{1}{}{H_{\text{KS}}}\prescript{1}{}{D})+w\text{Tr}(\prescript{1}{}{F}-\prescript{1}{}{D})$\\
such that\\
$\prescript{1}{}{D} \succeq 0$\\
$\prescript{1}{}{Q} \succeq 0$\\
$\left(\begin{array}{cc}
^{1} I & ^{1} D
\\
 ^{1} D & ^{1} F
\end{array}\right) \succeq 0$\\
$\prescript{1}{}{D} + \prescript{1}{}{Q} = \prescript{1}{}{I}$\\
};
\end{comment}
\begin{comment}
\node (sdp) [process, right=of fock] {$
\begin{aligned}
asdf=asdf
\end{aligned}
$};
\end{comment}
\node (converged?) [decision, below left=of sdp] {$\Delta$E $<$ tol?};
\node (converged) [startstop, left=of converged?] {Converged};
%\node (rdm1) [io, left=of converged?] {Build $\prescript{1}{}{D}$};
\draw [arrow] (start) -- (fock);
\draw [arrow] (fock) -- (sdp);
\draw [arrow] (sdp) -- (converged?);
\draw [arrow] (converged?) -- (converged);
\draw [arrow] (converged?) -- (fock);
\end{tikzpicture}
\caption{\textcolor{black}{Schematic description of the 1-RDMFT algorithm. From an initial guess density, $\prescript{1}{}{D}$, the Kohn-Sham 1-body Hamiltonian, ${H_{KS}[\prescript{1}{}{D}]}$, is generated using a traditional DFT exchange correlation functional. This Hamiltonian is then used in a SDP based minimization to yield a new 1-RDM. Self-consistent-field iterations are continued until the energy is converged below a selected target threshold.}}
\label{fig:flow}
\end{figure*}
%\end{comment}

\subsection{Comparison to iDMFT}\label{sec:cnnect}

We briefly review iDMFT~\cite{iDMFT} and then draw a comparison between iDMFT and our 1-RDMFT. The iDMFT postulates that the missing correlation energy from the Hartree-Fock method can be expressed as the fermionic entropy, defined as
\begin{equation}\label{eq:entropy}
    S = -\theta~\sum_{i}[n_{i}\ln{n_{i}}+(1-n_{i})\ln{(1-n_{i})}] \,,
\end{equation}
where $n_{i}$ is the occupation of the $i^{\rm th}$ orbital.  Minimizing the Hartree-Fock energy plus this entropic term leads to orbital occupations defined by the Fermi-Dirac distribution
\begin{equation}
    n_i=\frac{1}{1+\exp[(\epsilon_i-\mu)/\theta]}\,,
\end{equation}
where $\epsilon_{i}$ is the occupation of the $i^{\rm th}$ orbital, and $\mu$ is the chemical potential which is constrained such that $\sum_{i}n_{i}=N$.  This distribution naturally leads to non-idempotent 1-RDMs, with the degree of fractional occupation increasing as orbital energies become degenerate or the fictitious temperature $\theta$ increases. The fractional occupations are then used in the fermionic entropy correction to the energy which is simply subtracted from the total electronic energy of the system. iDMFT uses the same principles as thermally assisted occupation DFT (TAO-DFT) and the Fermi-smearing technique; however, it applies them to the Hartree-Fock theory instead of a density functional~\cite{Chai2012,Chai2014,mermin_1965}. In this work we have implemented iDMFT to utilize both the Hartree-Fock method (or functional) as well as available density functionals to facilitate a more thorough comparison with 1-RDMFT.\\

The energy correction of iDMFT is based on the information entropy of the 1-RDM while the correction of the SDP-based 1-RDMFT is based on the idempotency relation of the 1-RDM. These two corrections, we can show, agree with each other through second order in a Taylor series expansion of the information entropy. We first recast Eq.~(\ref{eq:entropy}) in terms of particle and hole matrices
\begin{equation}\label{eq:DQ_entropy}
    S = -\theta~\text{Tr}[\prescript{1}{}{D}\ln{(\prescript{1}{}{D})}+\prescript{1}{}{Q}\ln{(\prescript{1}{}{Q})}]\,,
\end{equation}
Second, we expand the natural logarithm of $\prescript{1}{}{D}$ in powers of $\prescript{1}{}{Q}$ through second order
\begin{align}
    \ln{(\prescript{1}{}{D})} &= \ln{(\prescript{1}{}{I}-\prescript{1}{}{Q})}\,,\\
    &=-\sum_{n=1}^{\infty}\frac{\prescript{1}{}{Q}^n}{n}\,,\\
    &=-\prescript{1}{}{Q}-\frac{\prescript{1}{}{Q}^{2}}{2}-O(\prescript{1}{}{Q}^3)\,. \label{eq:lnd}
\end{align}
Similarly, by particle-hole symmetry
\begin{equation}
\label{eq:lnq}
    \ln{(\prescript{1}{}{Q})} = -\prescript{1}{}{D}-\frac{\prescript{1}{}{D}^{2}}{2}-O(\prescript{1}{}{D}^3)\,.
\end{equation}
Inserting Eqs.~(\ref{eq:lnd}) and~(\ref{eq:lnq}) into Eq.~(\ref{eq:DQ_entropy}) yields
\begin{equation}
   \label{eq:Sapp}
    S \approx -\theta~\text{Tr}[\prescript{1}{}{D}(-\prescript{1}{}{Q}-\frac{\prescript{1}{}{Q}^{2}}{2})+\prescript{1}{}{Q}(-\prescript{1}{}{D}-\frac{\prescript{1}{}{D}^{2}}{2})]\,.
\end{equation}
Upon substituting the identity $^{1} Q = ^{1} I - ^{1} {D}$ and simplifying without further approximation, we obtain
\begin{equation}\label{eq:2nd_order}
    S \approx -\frac{5}{2}~\theta~\text{Tr}[\prescript{1}{}{D}^{2}-\prescript{1}{}{D}]\,.
\end{equation}
Comparing this form with Eq.~(\ref{eq:fin_correction}) reveals that our energy correction and iDMFT's correction agree with each other through second order in the expansion of the natural logarithms of the 1-particle and 1-hole RDMs in the information entropy for $\theta = 2w/5$. Furthermore, it can also be seen from Eq.~(\ref{eq:Sapp}) that truncation of the logarithmic expansions to only first order also yields Eq.~(\ref{eq:2nd_order}) with the scalar value of 5/2 replaced by 2. \\

\section{Results}
In this work we utilize our 1-RDMFT method with a simple weight matrix defined as ($\prescript{1}{}{W} = w \, ^{1} I$), where $w$ is \textcolor{black}{a} system specific \textcolor{black}{constant} and is optimized to reproduce either the dissociation energy or rotational barrier obtained from full CI (FCI) or the anti-Hermitian contracted Schr\"odinger equation (ACSE)~\cite{FCI1,FCI2,ACSE1,ACSE2,ACSE3} calculations. Three different functionals of varying HF exchange are surveyed and reported in the text, namely SCAN~\cite{SCAN}, M06-2X~\cite{M06-2X}, and HF with percent HF exchanges of 0\%, 54\% and 100\%, respectively. Additional data obtained with the functionals M06-L~\cite{Truhlar2006}, B3LYP~\cite{BECKE-B3LYP, STEPHENS-B3LYP}, M06~\cite{Truhlar2006}, and M06-HF~\cite{M06-HF} are reported in \textcolor{black}{Tables S1 and S2} in the SI. All DFT, HF and 1-RDMFT calculations were performed using PySCF~\cite{PySCF} with the cc-pVDZ basis set~\cite{ccpvdz}, while the ACSE calculations \textcolor{black}{for CO, N2, HF and C2H} were performed in Maple using the Quantum Chemistry Toolbox~\cite{Maple,QCP}. \textcolor{black}{All 1-RDMFT calculations were started with the converged 1-RDM from the respective DFT or HF results.} The SDP was solved using a boundary-point SDP algorithm previously developed by one of the authors for solving the variational 2-RDM problem~\cite{SDP2,Schlimgen2016}.\\

\begin{figure}[ht!]
    \includegraphics[width=.8\linewidth]{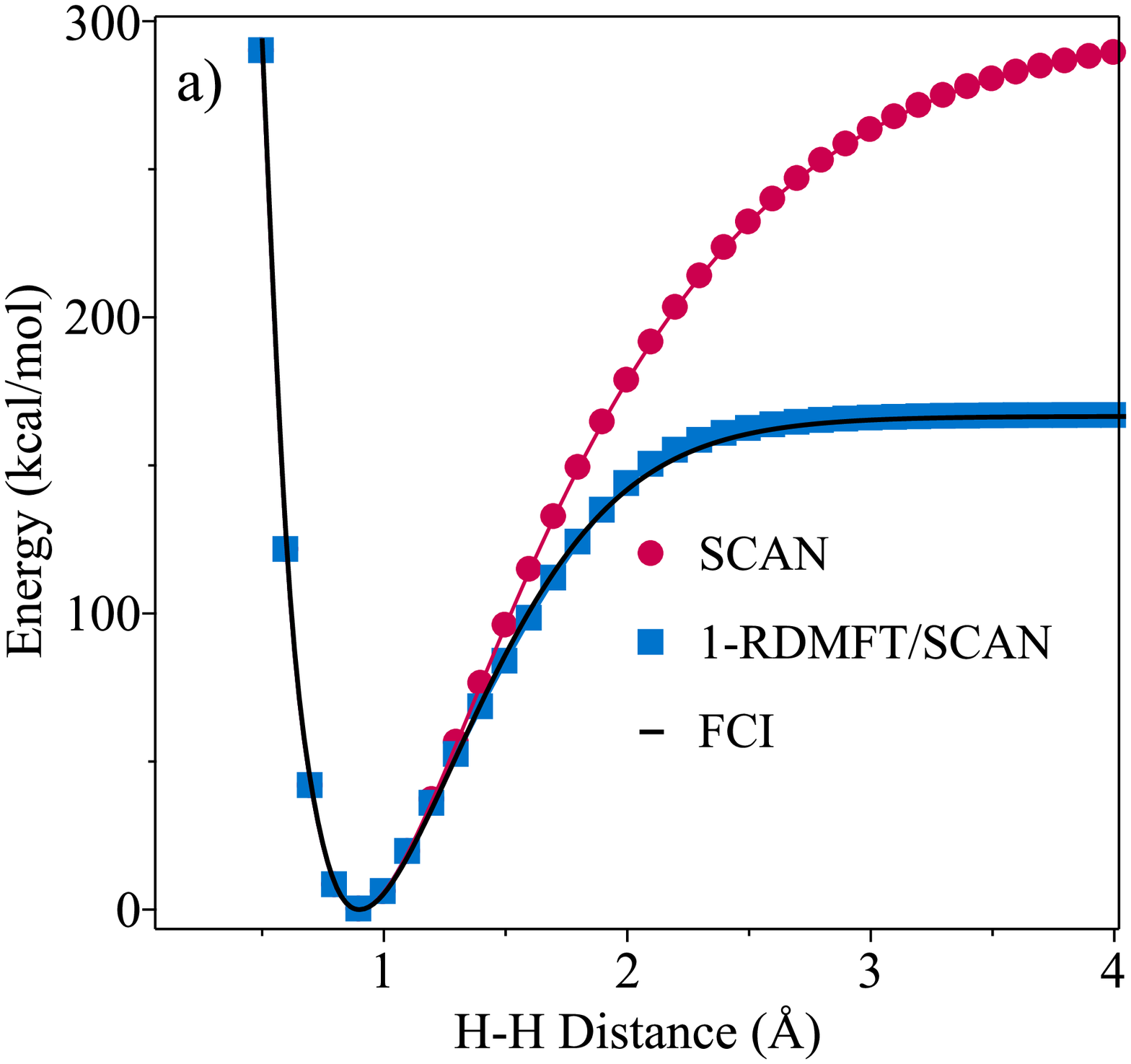}
    \includegraphics[width=.8\linewidth]{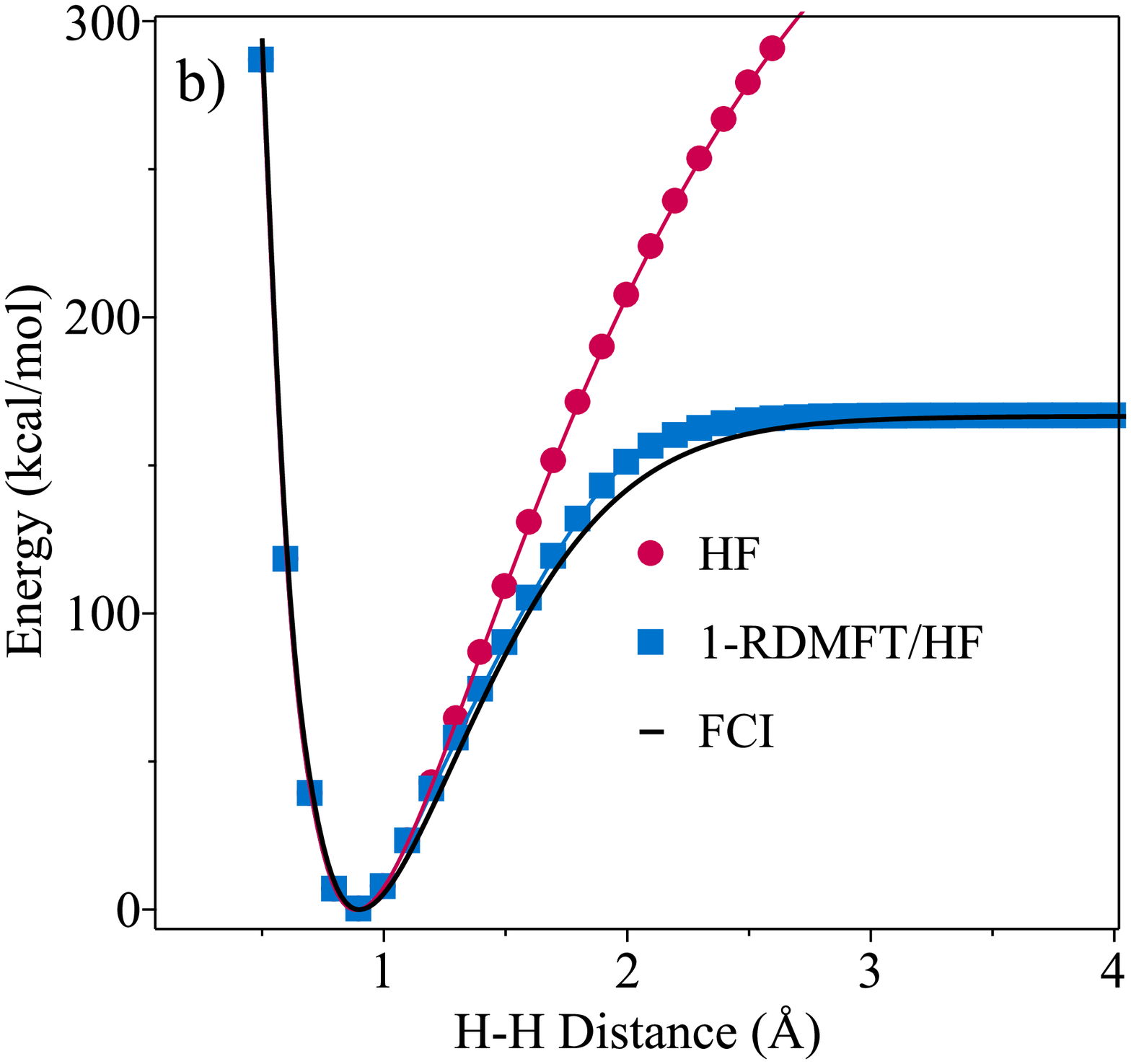}
\caption{Plot of the linear dissociation of H$_4$ in the cc-pvdz basis set with \textcolor{black}{equal distances} between all pairs of \textcolor{black}{adjacent} \textcolor{black}{hydrogens}. a): Comparison of the SCAN functional in the traditional KS-DFT implementation and within our 1-RDMFT method using a $w$ value of 0.104 to FCI. b): Comparison of HF in its traditional formulation and within our 1-RDMFT method using a $w$ value of 0.249 to FCI.}
\label{fig:H4}
\end{figure}

While both KS-DFT and HF struggle with the capture of static correlation due to their single-reference nature, KS-DFT's errors are generally smaller than those produced by HF~\cite{Baerends2003}. This is attributable to KS-DFT's approximate exchange correlation functional that captures some correlation effects.  To investigate, we apply conventional DFT and our 1-RDMFT with the the SCAN and HF functionals to~\textcolor{black}{linear} H$_{4}$ \textcolor{black}{where all adjacent hydrogens are equally spaced}, displayed in Figure \ref{fig:H4}.  The data reveals that in the KS-DFT formalism the SCAN functional fails to describe the dissociation limit, yielding an error of 122.72 kcal/mol at 4~\AA while in 1-RDMFT method it accurately reproduces the dissociation curve including the dissociation limit. \textcolor{black}{Plots of the errors in the potential energy curves are shown in Figures S1 and S2 in the SI.}  Furthermore, although the HF method yields an even greater error of~218.68 kcal/mol at 4~\AA{} than the DFT functional SCAN, it also yields an accurate potential energy curve in the framework of the 1-RDMFT.  Minor differences arise in the stretched bonding region of 1.8-2.2~\AA{} where the HF-based 1-RDMFT overestimates the energy, which results in a faster approach to the dissociation energy limit relative \textcolor{black}{to} the \textcolor{black}{FCI} reference curve. \\

\begin{figure}[ht!]
    \includegraphics[width=.8\linewidth]{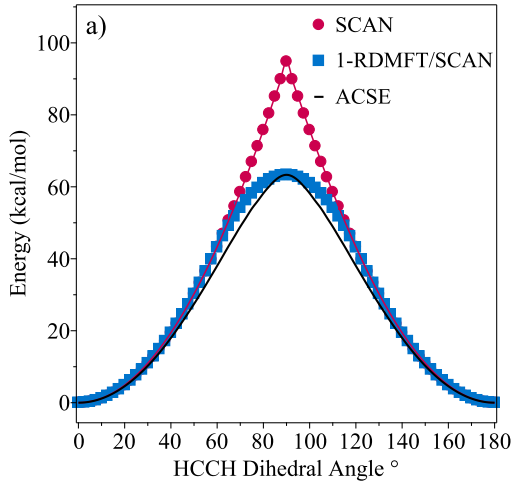}
    \includegraphics[width=.8\linewidth]{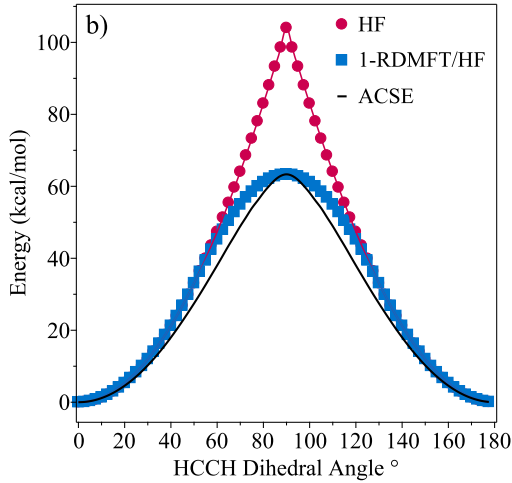}
\caption{Plot of the rotation of C$_2$H$_4$ along its HCCH dihedral angle. The relative energies are zeroed to the planar geometry at 0$\degree$. a): Comparison of the SCAN functional in the traditional KS-DFT implementation and within our 1-RDMFT method using a $w$ value of 0.052 to the ACSE. b): Comparison of HF in its traditional formulation and within our 1-RDMFT method using a $w$ value of 0.215 to the ACSE.}
\label{fig:ethene}
\end{figure}

\begin{table*}[ht]
    \centering
    \caption{1-RDMFT \textcolor{black}{and iDMFT error} values \textcolor{black}{used to quantify their reproduction of the dissociations of H$_2$, H$_4$, N$_2$, HF, and CO as well as the bond rotation of C$_2$H$_4$}. The maximal errors are defined as E$_{\textcolor{black}{\text{1-RDMFT/iDMFT}}}$ - E$_{\textcolor{black}{\text{FCI/ACSE}}}$ with the largest absolute magnitude being selected. Reference energies are computed from FCI for H$_{2}$ and H$_{4}$ and from ACSE for the other molecules.  The signed and unsigned errors are obtained as the average deviation from the reference curve from equilibrium to 4~\AA{} using 0.1~\AA{} step sizes.}
    \begin{ruledtabular}
    \begin{tabular}{crrrrrrrrrr}
                &  & \multicolumn{3}{c}{Maximal Error kcal/mol} & \multicolumn{3}{c}{Signed Error \textcolor{black}{over PES}} & \multicolumn{3}{c}{Unsigned Error \textcolor{black}{over PES}}\\\cline{3-5}\cline{6-8}\cline{9-11}
        Method  & & SCAN   & M06-2X&   HF      &  SCAN  & M06-2X&  HF  &  SCAN & M06-2X&  HF\\\hline
        1-RDMFT & H$_2$   & ~6.14  & 11.53 & 14.11     & 1.71  & 2.60  & 4.19  & 1.71  & 2.64 & 4.19\\
                & H$_4$   & -2.88  & ~2.66 & ~9.02     & 0.34  & 0.40  & 3.24  & 1.00  & 0.98 & 3.24\\
                & N$_2$   & ~9.99  & 19.01 & 34.88     & 0.87  & 4.22  & 11.20 & 1.65  & 4.24 & 11.20\\
                & HF      & -2.35  & ~6.87 & 17.39     & -1.29 & 0.30 & -0.39  & 1.30  & 2.05 & 7.06\\
                & CO      & ~3.80  & 17.31 & 35.05     & 0.79  & 4.99  & 9.84  & 1.68  & 5.89 & 11.06\\
                & C$_2$H$_4$  & ~5.54  & ~4.98 & ~7.21 & 2.07  & 1.76  & 3.10  & 2.07  & 1.76 & 3.10\\
                & RMSE     & \textcolor{black}{5.72}   & \textcolor{black}{12.06} & \textcolor{black}{22.66}	   & \textcolor{black}{1.31}  & \textcolor{black}{2.97}  & \textcolor{black}{6.58}  & \textcolor{black}{1.60}  & \textcolor{black}{3.36} & \textcolor{black}{7.48}\\\cline{2-11}
        iDMFT   & H$_2$  	   & ~-0.44 & ~-1.88 & -1.22 & -0.12 & -0.64 & -0.36 & 0.15 & 0.67 & 0.37\\
                & H$_4$	   & ~-8.38 & -12.01 & -9.87 & -2.80 & -4.59 & -3.56 & 2.81 & 4.59 & 3.57\\
                & N$_2$	   & -13.90 & -12.03 & ~2.92 & -4.33 & -3.97 &  0.51 & 4.33 & 3.97 & 0.85\\
                & HF	       & ~-4.79	& ~-4.08 & ~3.56 & -2.07 & -2.32 & -0.91 & 2.11 & 2.40 & 2.23\\
                & CO	       & ~-7.89	& ~-5.78 & -8.50 & -2.62 & -2.10 & -0.40 & 2.77 & 2.30 & 3.05\\
                & C$_2$H$_4$ & ~~2.28	& ~-2.24 & -2.40 & 0.97  & -1.04 & -1.03 & 0.97 & 1.04 & 1.05\\
                & RMSE	       & \textcolor{black}{7.68}	& \textcolor{black}{7.61}	 & \textcolor{black}{5.75}	 & \textcolor{black}{2.54}	 & \textcolor{black}{2.83}	 & \textcolor{black}{1.59}	 & \textcolor{black}{2.57} & \textcolor{black}{2.87} & \textcolor{black}{2.20}\\
    \end{tabular}
    \end{ruledtabular}
    \label{tab:iDMFT}
\end{table*}

Next, we consider the C-C bond rotation in C$_2$H$_4$. Here, the SCAN functional in KS-DFT overestimates the barrier height by 31 kcal/mol while in 1-RDMFT the barrier height \textcolor{black}{can be recovered within} sub milli-kcal/mol accuracy (Figure \ref{fig:ethene}). Although the 1-RDMFT curve is unable to match the reference with the same level of accuracy as seen for H$_4$, it is able to remove the non-physical discontinuity observed from KS-DFT at the 90\degree~ dihedral angle. This discontinuity is attributable to the increasingly diradical nature of the molecule as the dihedral angle approaches 90\degree~ and the highest occupied natural orbital (HONO) and lowest unoccupied natural orbital (LUNO) become degenerate, which may not be properly described using a single Slater determinant. Using HF in its traditional implementation also results in a non-physical discontinuity, as well as a barrier height error of 41 kcal/mol, which is an increase relative to the barrier height error of 31 kcal/mol from SCAN.  However, using HF within our 1-RDMFT framework \textcolor{black}{we} once again reproduce the barrier height to within sub milli-kcal/mol accuracy although the deviations from the reference curve are, as with H$_4$, larger than those seen using SCAN. Furthermore, while KS-DFT deviates from the 1-RDMFT curve when using SCAN at a dihedral angle of 57.5\degree, traditional HF deviates from the 1-RDMFT curve when using HF at a smaller angle of 50\degree. This points towards the 1-RDMFT correction resulting in non-idempotency earlier in the bond rotation for HF than SCAN as our 1-RDMFT method produces no energetic change until idempotency is broken. \textcolor{black}{This is verified through investigations of the non-idempotent residual in Figures S5 and S6 in the SI.}\\

Expanding our test set to also include the statically correlated dissociations of H$_2$, N$_2$, HF and CO, as well H$_4$ and C$_2$H$_4$, we tabulate the maximal deviation and average signed and unsigned errors \textcolor{black}{from the reference potential energy curve (PEC)} in Table~\ref{tab:iDMFT}. The reference PEC is computed from FCI for H$_{2}$ and H$_{4}$ and from ACSE for the other molecules; the signed and unsigned errors for the dissociations are computed from the average errors in the reference curve between equilibrium to 4 \AA{} in 0.1 \AA{} step sizes while for C$_2$H$_4$ the errors are calculated from finely spaced points over the entire dihedral angle. Comparing the surveyed density functionals in Table~\ref{tab:iDMFT} reveals that using any of them within the 1-RDMFT framework reproduces the reference curve more accurately than when using HF in 1-RDMFT, which consistently gives the largest errors. Comparing the \textcolor{black}{maximal errors of the} two density functionals SCAN and M06-2X in Table~\ref{tab:iDMFT} shows the SCAN functional obtaining a \textcolor{black}{root mean squared error (RMSE) of 5.72 kcal/mol}, approximately half that of M06-2X's \textcolor{black}{12.06} kcal/mol, with SCAN only having a slightly larger absolute error in two systems, H$_4$ and C$_2$H$_4$. Additionally, when including the functionals from \textcolor{black}{Tables S1 and S2 in the} SI which, in terms of increasing HF exchange, have \textcolor{black}{RMSEs of the maximal errors} of \textcolor{black}{6.43, 6.56, 9.46, 23.48} kcal/mol for M06-L, B3LYP, M06, and M06-HF respectively, it is observed that as the fraction of HF exchange in the DFT functional increases, the errors tend to increase as well. The \textcolor{black}{RMSEs of the mean signed errors being} \textcolor{black}{1.31, 2.97, 6.58} kcal/mol for SCAN, M06-2X, HF further clarifies that as the HF exchange increases, 1-RDMFT increasingly overestimates the energy of the system.  The relative improvements of the density functionals over HF is attributable to their recovery of additional dynamical correlation as the 1-RDMFT method is intended to treat strongly correlated orbitals, neglecting those with only small fractional occupations. The maximal errors consistently occur in the stretched bond region with both the equilibrium region and the dissociation limit being well described. There was not a noticeable trend between the fraction of HF exchange and the location of the maximal error along a curve.\\

\begin{table*}[]
    \caption{Optimized 1-RDMFT $w$ and iDMFT $\theta$ values and their ratios are reported.}
    \begin{ruledtabular}
    \begin{tabular}{cccccccccc}
                   & \multicolumn{3}{c}{SCAN} & \multicolumn{3}{c}{M06-2X} & \multicolumn{3}{c}{HF}\\\cline{2-4}\cline{5-7}\cline{8-10}
                   &  $w$  & $\theta$ & $w/\theta$ ratio &  $w$  & $\theta$ & $w/\theta$ ratio &  $w$  & $\theta$ & $w/\theta$ ratio \\\hline
        H$_2$  	   & 0.098 & 0.036 & 2.76 & 0.171 & 0.062 & 2.74 & 0.256 & 0.095 & 2.70\\
        H$_4$	   & 0.104 & 0.038 & 2.72 & 0.170 & 0.063 & 2.69 & 0.249 & 0.094 & 2.66\\
        N$_2$	   & 0.114 & 0.042 & 2.72 & 0.213 & 0.080 & 2.66 & 0.325 & 0.124 & 2.62\\
        HF	       & 0.073 & 0.024 & 3.11 & 0.189 & 0.061 & 3.08 & 0.318 & 0.105 & 3.04\\
        CO	       & 0.076 & 0.027 & 2.86 & 0.176 & 0.062 & 2.83 & 0.287 & 0.103 & 2.79\\
        C$_2$H$_4$ & 0.052 & 0.019 & 2.75 & 0.127 & 0.048 & 2.67 & 0.215 & 0.083 & 2.58\\
    \end{tabular}
    \end{ruledtabular}
    \label{tab:ratio}
\end{table*}

In the present exploratory calculations because $w$ is optimized for collections of calculations such as a given molecule's bond dissociation curve, we can use it as a gauge for the strength of the correction required for a given functional to capture static correlation. Comparing the $w$ values used for H$_4$ in Table~\ref{tab:ratio}, for example, gives 0.098, 0.171, 0.256 for the SCAN, M06-2X, and HF functionals, respectively. The data highlights that increasing HF exchange yields less accurate descriptions of statically correlated systems and therefore requires a larger correction weight $w$. The trend, not limited to H$_4$ but observed in all systems investigated here, supports previous work arguing that larger contributions of HF exchange in the exchange correlation functional increases static correlation errors within DFT~\cite{Truhlar2020}.\\

After investigating the use of different functionals within our 1-RDMFT framework, we compare it to the iDMFT method. For this comparison we utilize the same set of functionals and chemical systems, optimizing the adjustable parameter $\theta$ in iDMFT to reproduce the reference data. Starting with the maximum error from the reference curves, given in Table~\ref{tab:iDMFT}, it is evident that iDMFT reverses the trends seen in 1-RDMFT.  In iDMFT, instead of the local functional SCAN having the lowest maximum errors and HF having the largest, HF consistently gives lower errors relative to the DFT functionals. This is attributable to the over-inclusion of dynamical correlation which is treated in iDMFT through the density functional as well as the small non-zero orbital occupations in the entropic correction.  While 1-RDMFT and iDMFT have different trends in their maximum errors, the \textcolor{black}{RMSEs of the maximal errors} of the best performing functionals, SCAN and HF, are 5.\textcolor{black}{72}~kcal/mol and \textcolor{black}{5.75}~kcal/mol for 1-RDMFT and iDMFT respectively, leading to results of comparable accuracy between the two methods. Further support for their comparable accuracy is found in the \textcolor{black}{RMSE of their unsigned errors} where SCAN in 1-RDMFT has a \textcolor{black}{RMSE} of \textcolor{black}{1.60}~kcal/mol while HF in iDMFT has a \textcolor{black}{RMSE} of \textcolor{black}{2.20}~kcal/mol. Additionally, while H$_4$ displays the smallest deviations from the reference FCI curve using 1-RDMFT, H$_2$ is the system most accurately reproduced by iDMFT with a \textcolor{black}{signed error} of -0.12, -0.64, and -0.36 kcal/mol for SCAN, M06-2X, and HF respectively. Lastly, comparing the sign\textcolor{black}{s} of the  \textcolor{black}{signed errors} between the two methods, it is evident that while 1-RDMFT generally overestimates the energy along the curve, iDMFT typically underestimates it due to the over-inclusion of the dynamical correlation.\\

Because the adjustable parameter $\theta$ in iDMFT---like the $w$ in 1-RDMFT---is optimized for collections of calculations such as a given molecule's bond dissociation curve, its magnitude can again be used to gauge the degree of correction required for the HF and DFT functionals.  These $\theta$ values, given in Table~\ref{tab:ratio}, display the same trend observed in 1-RDMFT's $w$ values: as the fraction of HF exchange increases in the functional, the magnitude of $\theta$, reflecting the size of the correlation correction, increases. While the optimized values of $\theta$ are lower than the $w$ values, there is a consistent ratio between them with $w$ being {\raise.17ex\hbox{$\scriptstyle\mathtt{\sim}$}}2.7 times larger than $\theta$ (shown in Table \ref{tab:ratio}). Additionally, the ratio also appears to be affected by the amount of HF exchange in the functional, with it decreasing as the fraction of HF exchange increases. The computed ratios are in good agreement with the factor of 2.5 predicted from the Taylor series expansion in Section~\ref{sec:cnnect}. \\

\section{Conclusion}
In this work we have expanded on the theoretical underpinnings of our recently developed methodology which formally transforms traditional KS-DFT into a 1-RDFT via the inclusion of a 1-RDM based correction. A modified Kohn-Sham formalism, solvable by  semidefinite programming, allows for the accurate capture of strong correlation at favorable computational scaling.  Here we extend our 1-RDMFT to utilize the Hartree-Fock functional and delineate and formally derive the relation of our approach to the recently developed iDMFT method which introduces multi-reference correlation effects to HF via the use of an entropic correction, demonstrating the two theories are in agreement through second order.  We also extend iDMFT to use DFT functionals for better comparison to our 1-RDMFT.\\

To demonstrate the potentially broad applicability of our 1-RDMFT as well as to investigate its dependence on the chosen DFT functional, we have calculated the potential energy surfaces for several bond dissociations and the bond rotation of ethene, covering a range of different chemical bonding environments and functionals.  For the purpose of comparing HF and the DFT functionals, we optimize the $w$ parameter in 1-RDMFT (or $\theta$ parameter iDMFT) for each functional for a collection of calculations such as a molecule's potential energy curve. The results reveal that the 1-RDMFT can be effective at capturing multi-reference correlation across entire potential energy surfaces, smoothly interpolating between the single-reference equilibrium regime and the strongly correlated dissociated regime. Furthermore, we analyze the differences in the results from 1-RDMFT and iDMFT, obtaining a ratio between the $w$ and $\theta$ values of 1-RDMFT and iDMFT, respectively, of 2.7 that is in good agreement with the theoretically predicted value of 2.5.\\

While the general effectiveness of our weight-matrix correction, $w\prescript{1}{}{I}$, suggests that all functionals investigated in this work suffer from the same fundamental failings in describing static correlation---exemplified by the fact that only the adjustment of a scalar multiplier for each system and functional combination is required to obtain accurate surfaces---, the results also reveal fundamental differences in the various functionals' ability to capture strong correlation, quantified by the magnitude of the required 1-RDM correction.  In particular, we observe that the magnitude of the scalar value $w$ (or $\theta$ for iDMFT) depends on the amount of HF exchange included in a chosen density functional, with an increasing percentage of HF exchange requiring a larger $w$ and, hence, a bigger correction.  Interestingly, while a pure functional yields the best agreement with high-level reference data in our 1-RDMFT framework, the opposite is true in the case of iDMFT, which performs best with the HF method.\\

Future work will focus on the determination of the system-specific weight matrix, $w$, with particular promise being held by the use of machine learning\cite{Schmidt2019,Brockherde2017,Moreno2020,Margraf2021,Kalita2021} for this purpose.  More generally, the transformation of DFT into a 1-RDMFT presents a fresh paradigm for the prediction of both dynamic and static correlation at a mean-field-scaling computational cost.  Unlike traditional 1-RDMFT approaches, the present theory allows us to achieve a lower computational scaling by exploiting DFT's existing functionals, and in contrast to DFT, it allows us to harness the additional information of the 1-RDM, especially its fractional eigenvalues (natural-orbital occupations), to realize a more accurate description of static correlation, which has important applications to many molecular structures and processes.

\begin{acknowledgments}
D.A.M. gratefully acknowledges the U.S. National Science Foundation Grants No. CHE-2155082 and No. CHE-2035876.
\end{acknowledgments}

\section*{Supporting Information}
Additional functionals as well as energy deviations, convergence rates, non-idempotency vs HOMO-LUMO gaps, and Fermi-smearing figures can be found in the SI.

\bibliographystyle{achemso}

\bibliography{citations.bib}

\newpage

\begin{figure}
    \includegraphics[scale=0.25]{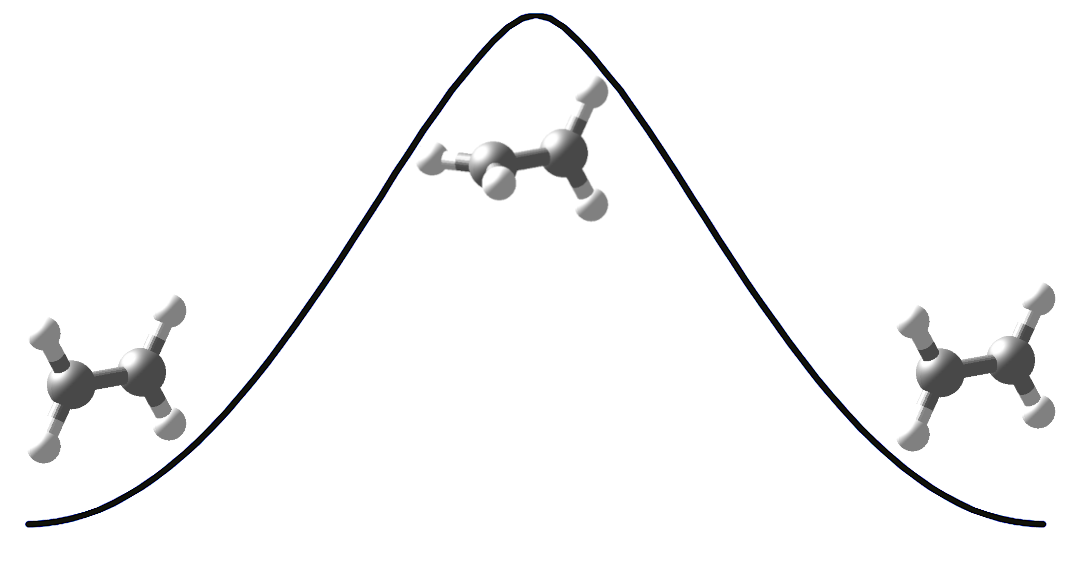}
    \caption{TOC}
\end{figure}

\end{document}